\documentstyle[preprint,aps,psfig]{revtex}
\newcommand{\beq}{\begin{equation}}
\newcommand{\eeq}{\end{equation}}
\newcommand{\beqa}{\begin{eqnarray}}
\newcommand{\eeqa}{\end{eqnarray}}
\newcommand{\ba}{\begin{array}}
\newcommand{\ea}{\end{array}}

\begin{document}

\draft
\widetext 

\title{Shell Effects and Phase Separation \\ 
in a Trapped Multi-Component Fermi System} 

\author{L. Salasnich and L. Reatto}

\address{Istituto Nazionale per la Fisica della Materia, 
Unit\`a di Milano Universit\`a, \\ 
Dipartimento di Fisica, Universit\`a di Milano, \\ 
Via Celoria 16, 20133 Milano, Italy}

\author{A. Parola}

\address{Istituto Nazionale per la Fisica della Materia, Unit\`a di Como, \\ 
Dipartimento di Scienze Fisiche, Universit\`a dell'Insubria, \\ 
Via Lucini 3, 23100 Como, Italy}

\maketitle

\begin{abstract}
Shell effects in the coordinate space 
can be seen with degenerate Fermi vapors 
in non-uniform trapping potentials. 
In particular, below the Fermi temperature, 
the density profile of a Fermi gas in a confining 
harmonic potential is characterized by several local maxima. 
This effect is enhanced for 
"magic numbers" of particles and in quasi-1D 
(cigar-shaped) configurations. 
In the case of a multi-component Fermi vapor, 
the separation of Fermi components in different 
spatial shells (phase-separation) depends on 
temperature, number of particles and scattering length. 
We derive analytical formulas, based on bifurcation theory, 
for the critical density of Fermions and the critical 
chemical potential, which give rise to the phase-separation.
\end{abstract}

\section{Introduction} 
 
The Fermi quantum degeneracy with trapped 
dilute vapors of $^{40}$K atoms has been experimentally 
achieved in 1999.\cite{p1} 
The s-wave scattering between Fermions in the same hyperfine state  
is inhibited due the Pauli principle. Moreover, for $^{40}$K atoms 
the p-wave scattering length becomes negligible 
in the limit of zero temperature. 
It follows that at very-low temperatures the dilute Fermi gas,  
in a fixed hyperfine state, is practically ideal. 
Nevertheless, the effect of interaction could be very effective  
for a Fermi vapor with two or more hyperfine states. 
In this paper we consider a Fermi vapor confined in 
a harmonic potential that models 
the trap of recent experiment.\cite{p1}
These trapped Fermi gases are quite interesting because 
the quantum degeneracy shows up not  
only in momentum space, as in uniform systems, but also  
in coordinate space. 
 
\section{Ideal Fermi gas at finite temperature} 
 
Let us consider an ideal Fermi gas in external potential 
described by the non-relativistic field 
\beq 
{\hat \psi}({\bf r})= \sum_{\alpha} \phi_{\alpha}({\bf r})
{\hat a}_{\alpha} \; , 
\eeq
where $\phi_{\alpha}({\bf r})=\langle{\bf r}|\alpha\rangle$ 
is the single-particle eigenfunction with eigenvalue 
$\epsilon_{\alpha}$ and ${\hat a}_{\alpha}$ is 
the lowering Fermi operator of the single-particle 
eigenstate $|\alpha \rangle$. 
The grand canonical thermal average of the Fermi spatial density 
${\hat \psi}^+({\bf r}){\hat \psi}({\bf r})$ is given by 
\beq
n({\bf r}) = \langle\langle {\hat \psi}^+({\bf r})
{\hat \psi}({\bf r}) \rangle\rangle = 
\sum_{\alpha} {|\phi_{\alpha}({\bf r})|^2  
\over e^{\beta(\epsilon_{\alpha}-\mu)} + 1}  \; ,  
\eeq  
where $\mu$ is the chemical potential and $\beta=1/(kT)$ 
with $k$ the Boltzmann constant and $T$ the absolute temperature.\cite{p2} 
Note that the thermal average of a generic operator ${\hat A}$ 
is defined as 
\beq
\langle\langle {\hat A}\rangle\rangle = 
{Tr\big[ {\hat A} \; e^{-\beta({\hat H}-\mu{\hat N})} \big] 
\over 
Tr\big[ e^{-\beta({\hat H}-\mu{\hat N})} \big] }  \; ,
\eeq
where ${\hat H}=\sum_{\alpha}\epsilon_{\alpha} 
{\hat a}_{\alpha}^+{\hat a}_{\alpha}$ 
is the Hamiltonian of the system and 
${\hat N}=\sum_{\alpha} {\hat a}_{\alpha}^+{\hat a}_{\alpha}$ 
is the number operator. The average number $N$ 
of particles of the system, given by 
\beq 
N=\sum_{\alpha} \langle\langle a_{\alpha}^+a_{\alpha} \rangle\rangle 
= \sum_{\alpha} {1\over e^{\beta(\epsilon_{\alpha}-\mu)} + 1} \; , 
\eeq 
fixes the chemical potential.\cite{p2} 
\par 
In the case of a harmonic external potential 
$U({\bf r}) = (m/2)(\omega_1^2 x^2 + \omega_2^2 y^2  
+\omega_3^2 z^2)$, one finds the Fermi density profile  
by using the Eq. (2) and 
the eigenfunctions $\phi_{n_1n_2n_3}({\bf r})$ 
of the harmonic oscillator  
\beq 
n({\bf r})= \sum_{n_1n_2n_3=0}^{\infty}  
{|\phi_{n_1n_2n_3}({\bf r})|^2  
\over e^{\beta\hbar(\omega_1(n_1+1/2)+\omega_2(n_2+1/2)  
+\omega_3(n_3+1/2)-\mu)} + 1} \; .  
\eeq  
Because the Fermi gas is ideal, one has 
$\phi_{n_1n_2n_3}({\bf r})=\phi_{n_1}(x) 
\phi_{n_2}(y)\phi_{n_3}(z)$, where $\phi_n(x)$ is the 
eigenfuction of a 1D harmonic oscillator with 
frequency $\omega$ and quantum number $n$. 
This eigenfunction can be found by means of 
the recursion relation\cite{p3}  
\beq 
\phi_{n}(x) = {1\over \sqrt{n}} \Big[ 
\sqrt{2} \sigma x \phi_{n-1}(x)- \sqrt{n-1} \phi_{n-2}(x) 
\Big] \; , 
\eeq
where $\phi_0(x)= \sigma^{1/2}\pi^{1/4} e^{-\sigma^2x^2/2}$ 
and $\phi_1(x)=\sqrt{2} \sigma x \phi_0(x)$, 
with $\sigma=(m\omega/\hbar)^{1/2}$.  

\begin{figure}
\centerline{\psfig{file=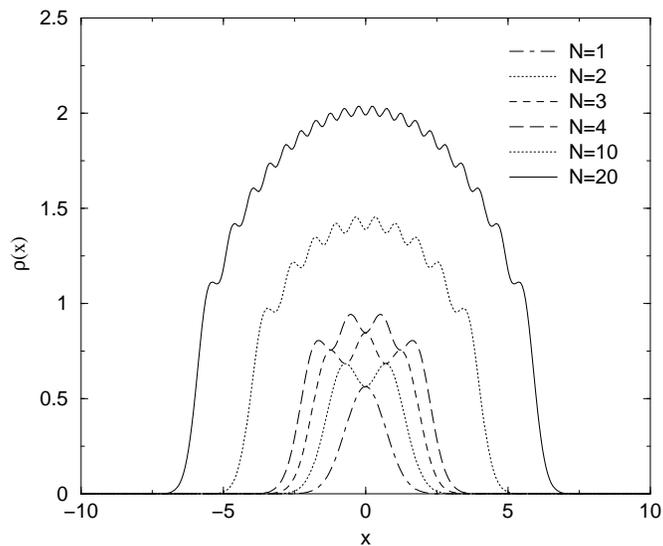,height=2.9in}} % postscript image 
\caption{Density profiles for an ideal Fermi gas in a 1D harmonic trap.
Temperature: $kT/\hbar \omega =10^{-3}$; at this 
temperature the results coincide with zero-temperature ones. 
$N$ is the number of Fermions. 
Lengths in units $a_H=(\hbar/m\bar{\omega})^{1/2}$  
and densities in units $a_H^{-3}$.}
\end{figure}

\par 
The Eq. (5) and (6) are used to numerically 
calculate the exact density profile. 
Such exact density profile can be compared with 
the semiclassical one, that is given by\cite{p4}  
\beq 
n({\bf r})={1\over \lambda^3} f_{3/2} 
\left(e^{\beta(\mu -U({\bf r}))}\right) \; , 
\eeq 
where $\lambda = (2\pi \hbar^2\beta /m)^{1/2}$ is the  
thermal length and  
\beq 
f_{n}(z) = {1\over \Gamma(n)} \int_0^{\infty} dx  
{x^{n-1}\over z^{-1}e^{x}+1} \; ,  
\eeq 
with $\Gamma(x)$ the factorial function. 

\begin{figure}
\centerline{\psfig{file=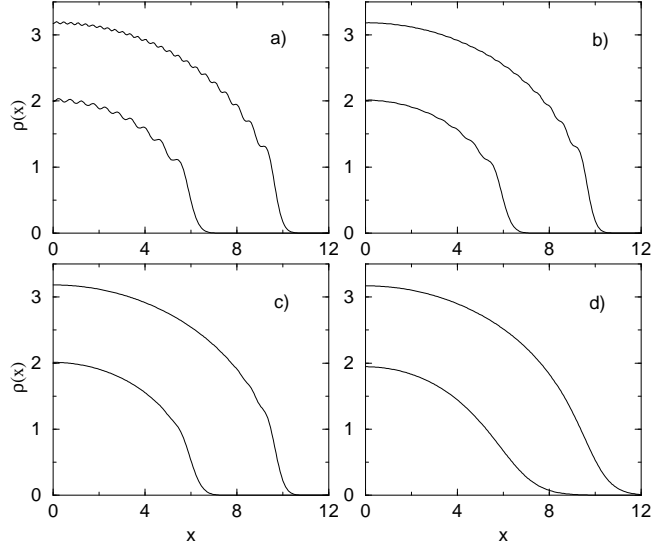,height=2.9in}} % postscript image 
\caption{Density profiles for an ideal Fermi gas in a 1D harmonic trap 
at finite temperature, with $N=20$ and $50$. 
$N$ is the number of Fermions. a) $kT/\hbar \omega = 1/10$; 
b) $kT/\hbar \omega = 1/2$; c) $kT/\hbar \omega = 1$; 
d) $kT/\hbar \omega = 5$. Units as in Figure 1.}
\end{figure} 

In the limit of zero temperature, with $\mu=E_F$ the Fermi energy, 
the semiclassical spatial distribution 
is given by the Thomas-Fermi approximation 
$n({\bf r})=(2m)^{3/2}/(6\pi^2\hbar^3) 
\left(E_F- U({\bf r})\right)^{3/2}  
\Theta\left(E_F- U({\bf r})\right)$, 
where $\Theta$ is the Heaviside step function. 
The Eq. (7) is the generalization of well-known formula  
for an ideal Fermi gas in a box that is 
exact in the thermodynamic limit.\cite{p2} 
Note that for $|z|<1$ one has $f_{n}(z)=  
\sum_{i=1}^{\infty} (-1)^{i+1} z^i/i^n$.  
Moreover, by using $g_n(z)=-f_n(-z)$ instead of  
$f_n(z)$, one finds the spatial distribution  
of the ideal Bose gas in external potential.\cite{p5,p6}  
The Fermi energy $E_F$ and the Fermi temperature $T_F$ 
are easily obtained by imposing the normalization 
condition to the Thomas-Fermi distribution. 
For a D-dimensional harmonic oscillator with a 
geometric average $\bar{\omega}=(\omega_1 ... \; \omega_D)^{1/D}$ 
of the $D$ frequencies, one finds the semiclassical result\cite{p7} 
\beq
E_F=kT_F = \left( D! N\right)^{1/D} \hbar \bar{\omega}  \; . 
\eeq 
For $D=3$ one recovers the familiar result 
$E_F=6^{1/3} N^{1/3} \hbar \bar{\omega}$.   
\par  
In a recent paper we have performed a 
comparison between exact and semiclassical results.\cite{p4} 
We have found that when $(kT/\hbar \bar{\omega})>1$, 
where $\bar{\omega}=(\omega_1\omega_2\omega_3)^{1/3}$, 
there are no appreciable deviations between exact and 
semiclassical results. Instead, for an isotropic harmonic potential, 
when $(kT/\hbar \bar{\omega})<1$ some differences are  
observable, in particular for "magic" numbers of particles  
($N=(p+1)(p+2)(p+3)/6$, where $p$ is a natural number) 
that correspond to a complete shell 
occupation of single-particle energy levels. 
The differences are reduced by increasing $N$ 
showing that semiclassical approximation provides an excellent 
representation of Fermi distribution for a wide range of 
parameters. In correspondence of the "magic numbers", the exact 
spatial density profile shows local maxima, 
which suggest a spatial shell structure. 
The magic numbers are particularly stable; 
in fact, for small variations of the chemical potential $\mu$ 
the magic number $N$ remains unchanged. 
The shell structure in the density profile 
is washed out by increasing the number of particles 
and is completely absent in the semiclassical approximation. 
\par 
In the quasi-1D case, namely a cigar-shaped 
gas where $\omega_1=\omega_2 >> \omega_3$, 
the shell effects are strongly enhanced. 
This can be shown by investigating the one-dimensional 
problem, as recently done by Kolomeiski 
et al\cite{p8} at $T=0$. We extend the 
calculations of Kolomeiski at finite temperature. 
In Fig. 1 we plot the density profile of 
a 1D ideal Fermi gas in harmonic potential 
as a function of the number $N$ of particles 
at $kT/\hbar \omega=10^{-3}$. 
We have verified that at this temperature the 
density profiles coincide with the zero-temperature ones. 
The results are obtained by numerically evaluating 
expressions (5) and (6) in the 1D case. 
The local maxima, whose number grows with $N$, are clearly visible 
for a small number of particles. 
In Fig. 2 we show the density profiles with $20$ and $50$ 
particles as a function of temperature. 
Remarkably, the local peaks are no more distinguishable 
for temperatures well below the Fermi temperature 
$T_F=N\hbar \omega/k$. Thus, to see spatial 
shell effect on the 1D density profile, the system 
should be at temperatures lower than $T_F$ 
by one or two orders of magnitude. 
 
\section{Phase separation with two Fermi components} 

In the recent experiment with dilute $^{40}$K 
atoms,\cite{p1} to favor the evaporative cooling, 
a $^{40}$K Fermi vapor in two hyperfine states  
($|9/2,9/2\rangle$ and $|9/2,7/2\rangle$) has been used. 
The problem of a dilute Fermi vapor with two 
hyperfine states (components) 
can be studied by using the s-wave scattering approximation, 
the mean-field approach and semiclassical formulas. 
The spatial density profiles of the two components 
of a Fermi vapor can be written as  
$$  
n_1({\bf r})={1\over \lambda^3}  
f_{3/2}\left(e^{\beta \left(\mu_1 -U({\bf r}) - g n_2({\bf r}) 
\right)}\right) \; ,  
$$  
\beq
n_2({\bf r})={1\over \lambda^3}  
f_{3/2}\left(e^{\beta \left(\mu_2 -U({\bf r}) - g 
n_1({\bf r}) \right)}\right) \; ,  
\eeq
where $\mu_i$ is the chemical potential of the  
$i$-th component, and $g=4\pi\hbar^2a_{12}/m$, with 
$a_{12}$ the s-wave scattering length between 
first and second component ($a_{11}=a_{22}=0$). 
Thus, the effect of the second Fermi component on  
the first component is the appearance of 
a mean-field effective potential.  
\par 
When two components have the same number of particles, the onset of  
phase-separation is also an example of spontaneous symmetry breaking.  
In fact, if the chemical potentials $\mu_i$ of the two components  
are equal, Eq. (10) always admits a symmetric solution  
$n({\bf r})=n_1({\bf r})=n_2({\bf r})$. However, for particle number $N$  
larger than a threshold $N_c$ the solution bifurcates and a pair of  
symmetry breaking solutions appears. Just beyond threshold the asymptotic  
solutions begin to differ from the symmetric one in  
a neighborhood of the origin ${\bf r}={\bf 0}$, i.e. at the point of higher  
density. An analytic formula for the critical chemical  
potential $\mu_c$ can be obtained by using the bifurcation theory.  
The system (10) can be re-written at the origin as 
\beq
{\bf F}^{\mu}({\bf x})={\bf 0} \; , 
\eeq
where 
$$
{\bf x}=\Big(x_1,x_2\Big)=\Big(n_1({\bf 0}),n_2({\bf 0})\Big)
$$ 
$$
{\bf F}^{\mu}({\bf x})=
\Big(F^{\mu}_1({\bf x}),F^{\mu}_2({\bf x})\Big)=
\Big(x_1-{1\over \lambda^3}
f_{3/2}\big(e^{\beta \left(\mu - g x_2) \right)}\big), 
x_2-{1\over \lambda^3}f_{3/2}
\big(e^{\beta \left(\mu - g x_1) \right)}\big)\Big) 
\; .
$$ 
For $\mu<\mu_c$ there exist only one symmetric solution 
of the previous equation, given by ${\bf x}^*=(x^*,x^*)$. 
At $\mu=\mu_c$ there are two solutions 
and the Jacobian of the function ${\bf F}^{\mu}({\bf x})$ 
must have the determinant equal to zero, i.e. 
\beq 
det\left[{\partial {\bf F}^{\mu}
\over \partial {\bf x}} ({\bf x}^*)\right]
= \left| \begin{array}{cc}  
         1 &  {1\over \lambda^3}{\partial\over \partial x_2}
f_{3/2}(e^{\beta\left(\mu - g x_2\right)})\\ 
{1\over \lambda^3}{\partial\over \partial x_1}
f_{3/2}(e^{\beta\left(\mu - g x_1\right)}) & 1  
\end{array}  \right|_{{\bf x}^*} =0 \; . 
\eeq 
By imposing this condition on Eq. (10) and (11) 
and setting $x^*=n({\bf 0})$, one finds 
\beq  
{g\beta \over \lambda^3} e^{\beta\left(\mu - g n({\bf 0})\right)}  
f_{3/2}'(e^{\beta\left(\mu - g n({\bf 0})\right)}) = 1 \; .  
\eeq  
At $T=0$, by the use of the first term of the large $z$ expansion  
of $f_{3/2}(z)$, given by\cite{p2} 
\beq
f_{3/2}(z)= {4\over 3\pi^{1/2}} 
\left[ \left(\ln{z}\right)^{3/2} + 
{\pi^2\over 8} \left(\ln{z}\right)^{-1/2} + ... \; \right] \; , 
\eeq 
one finds analytical expressions for the critical 
density $n_c({\bf 0})$ and the critical chemical 
potential $\mu_c$:  
\beq 
n_c({\bf 0})={\pi\over 48 a^3} \; , \;\;\;\;  
\mu_c={5 \pi^2 \over 24} \hbar\bar{\omega}  
\left({a_H\over a}\right)^2 \; .  
\eeq  
These remarkably simple formulas can be very useful  
to determine the onset of phase-separation in  
future experiments. Moreover, by knowing the critical chemical potential  
one numerically finds the number of particles. 
The conclusion is that  
by increasing the interaction between the two components  
one can use lower number of particles  
to obtain the phase-separation.  
At finite temperature one can use the first two terms 
of the large $z$ expansion of $f_{3/2}(z)$, i.e. Eq. (14). 
The analytical formulas at zero and finite temperature 
are in good agreement with the numerical calculations.\cite{p4} 

\section{Phase separation with many Fermi components} 

Phase-separation also appears in a Fermi vapor  
with three or more components.  
In Ref. 4 we have numerically solved 
the three-component extension of Eq. (10) 
with $a_{12}=a_{13}=a_{23}=a$. Also for three components 
the spontaneous symmetry breaking and the 
phase-separation are controlled by scattering length, 
temperature and total number of particles. 
In particular, one finds that by increasing the scattering length 
at first one of the components separates from the others, 
which remain still mixed. The separation begins at 
the center of the trap and, as the scattering length is further 
increased, also the other two components separate 
(this second phase-separation begins at the interface 
with the previously separated component) and 
one eventually sees complete phase-separation and 
the formation of $4$ or $5$ shells. 
\par 
The Eqs. (15) can be extended to a M-component Fermi 
vapor with the same number of particles in each component.  
The critical density $n_c({\bf 0})$ does not depend on the 
number $M$ of Fermi components and one gets the same result of 
Eq. (15). Instead, the critical chemical potential reads 
$\mu_c=(2M+1)\pi^2 \hbar \bar{\omega} (a_H/a)^2/24$. 
  
\section{Conclusions}  

We have analyzed a degenerate 
Fermi gas in a harmonic external potential. 
Below the Fermi temperature $T_F$ 
and for a small number of particles, shell effects, 
like local maxima in the spatial density profile 
for "magic numbers", are clearly visible. 
Such effects are strongly enhanced in the 1D case, 
but they are visible only for temperatures which are 
at least one order lower than $T_F$, 
i.e. far from the semiclassical limit. 
\par  
We have also considered a Fermi vapor with many hyperfine states. 
By using the bifurcation theory, 
we have shown that the onset of phase-separation appears  
by increasing the scattering length or, for a fixed 
scattering length, by increasing the number of particles.  
By raising the temperature, a larger scattering length or  
a larger number of particles is needed to obtain  
the phase-separation. A Fermi vapor with three or more 
components has the same behavior but at first only one of the 
components separates from the others, which remain still mixed. 
The critical density of Fermions 
at the origin, which gives rise to the phase-separation,   
satisfies the equation $n_c({\bf 0})=\pi/(48 a^3)$,  
where $a$ is the s-wave scattering length.


\begin{thebibliography}{99}
 
\bibitem{p1} B. DeMarco and D.S. Jin, {\it Science} {\bf 285}, 
1703 (1999).  
 
\bibitem{p2} K. Huang, 
{\it Statistical Mechanics} (Wiley, New York, 1980). 

\bibitem{p3} C. Cohen-Tannouji, B. Diu and F. Lalo\"e, 
{\it Quantum Mechanics} (Wiley, New York, 1987).  

\bibitem{p4} L. Salasnich, B. Pozzi, A. Parola and 
L. Reatto, J. Phys. B {\bf 33}, 3943 (2000). 

\bibitem{p5} B. Pozzi, L. Salasnich, A. Parola, L. Reatto,  
J. Low Temp. Phys. {\bf 119}, 57 (2000). 

\bibitem{p6} B. Pozzi, L. Salasnich, A. Parola, L. Reatto, 
Eur. Phys. Jour. D {\bf 11}, 367 (2000).  

\bibitem{p7} L. Salasnich, e-preprint math-ph/0008030, 
to be published in J. Math. Phys. {\bf 41} (2000). 

\bibitem{p8} E.B. Kolomeisky, T.J. Newman, J.P. Straley 
and X. Qi, Phys. Rev. Lett. {\bf 85}, 1146 (2000). 
 
\end{thebibliography}
\end{document}